\documentstyle[aps,prl,manuscript]{revtex} 
\begin{document}             

\draft

\title{Rotation in an asymmetric multidimensional periodic potential due
to colored noise}
\bigskip

\author{Avik W. Ghosh$^{*}$ and Sanjay V. Khare$^{\dagger}$}

\address{Department of Physics, Ohio State University, 174 West 18th Avenue,
Columbus, OH 43210} %

\maketitle
\medskip

\widetext
\begin{abstract}
We analyze the motion of an overdamped classical particle in a
multidimensional periodic potential, driven by a weak external noise. We
demonstrate that in steady-state, the presence of temporal correlations in the 
noise and spatial asymmetry within a period of the potential could lead to
particle rotation. The rotation is a direct consequence of a change in sign of 
the noise-induced drift motion in each dimension. By choosing different potentials, 
we can generate a variety of flow patterns from laminar drifts to rotations.

\end{abstract}
\bigskip

%

Recently there has been intense activity in the analysis of
stochastically driven ratchets \onlinecite{rAst3,r2,r1,rAst,dhr94,rLeib,rAst2,rM}.
These ratchets are spatially periodic systems where a spatial
asymmetry in the potential imposes a directionality, while ``memory''
effects from a temporally-correlated stochastic force (colored noise) or 
explicit time-dependences in the potential itself break detailed
balance, thereby rectifying microscopic fluctuations to 
generate a unidirectional particle drift. This type of
unidirectional drift has been verified through experiments on colloidal 
particles or polystyrene spheres \onlinecite{r10,r102}, cold rubidium atoms in an
asymmetric optical lattice \onlinecite{r11} and quantum dots
\onlinecite{r12}. Predictions have also been made for
SQUIDs\onlinecite{r13}. The applications of these
concepts have been manifold: this mechanism has been proposed as a
possible explanation for the long-range cellular transport of motor
proteins, utilizing temporal fluctuations at a 
microscopic scale to effectively transduce chemical energy into directed
mechanical work \onlinecite{r3}. In addition, such 
ideas are of interest in the nanoscale fabrication of devices.

Most recent analyses of ratchets have been limited to one-dimensional
(or ``linear'') molecular motors \onlinecite{r3}. In this paper we show that 
for a particle in a multidimensional potential, a time-correlated noise can break detailed 
balance and generate rotations. Beyond their intrinsic interest and
novelty, such rotations could be used to design non-equilibrium  molecular 
engines.
To demonstrate rotations, we develop a Fokker-Planck equation in
arbitrary dimensions for a weak Gaussian random noise and explicitly
demonstrate the existence of rotation in the presence of spatial asymmetry
and temporal correlations. Our argument can be simply described as
follows: in the presence of spatial asymmetry in one-dimension, a
time-correlated noise is known to produce a drift \onlinecite{rAst3,r2,r1,rAst,dhr94,rLeib,rAst2,rM};  the direction of the
drift is determined by the sense of the potential asymmetry. For a
potential in multiple dimensions, the sense of the potential asymmetry
along one coordinate can be reversed by varying the other coordinates,
leading to a change in sign (or reversal of asymmetry) of the
potential. As indicated in Fig. \ref{f1}, the coordinate-dependent
reversal of the one-dimensional drifts could then conspire together to
generate a rotation. Contrary to rotation generated in a potential,
 which is determined by the
initial conditions of the particle, the sense of our rotation is given
{\it{entirely}} by the potential asymmetry, and is in fact
{\it{independent}} of any initial conditions.  Furthermore, removing either the potential 
asymmetry or the correlation in the noise destroys the rotation.

We put our general argument outlined above in a mathematical form as
follows. We start with a Langevin equation describing an overdamped
particle driven by a Gaussian distributed colored noise of 
{\it{arbitrary}} correlation function in an {\it arbitrary}
multi-dimensional periodic potential. For weak noise and small 
correlation time, we derive a Fokker-Planck equation for the probability
density of the particle. We then show that in steady-state, the curl of the current density cannot be zero
everywhere in space as long as the correlation time is non-zero and the
periodic potential is asymmetric. Finally we demonstrate that relaxing
either of these two conditions leads to zero current density.

The Langevin equation describing the dynamics of an overdamped particle in a 
multi-dimensional space ${\vec{r}}$ in the presence of a potential 
$U(\vec{r})$ and a noise $\vec{f}(t)$ is given by \onlinecite{r6}:

\begin{equation}
\dot{\vec{r}}(t) = \vec{W}(\vec{r}) + \vec{f}(t)
\label{e1}
\end{equation}
where dot represents a time-derivative and $\vec{W} =
-\vec{\nabla}U(\vec{r})$ is the force exerted by the potential
$U(\vec{r})$. The noise is assumed to be Gaussian distributed, with a
correlation time $\tau^c_i$ and strength $D_i$ along the $i$th
coordinate. Generalizing Fox's functional calculus approach
\onlinecite{r7}, we express the probability distribution $P(\vec{r})$
in terms of a functional integral over different realizations of the
noise with a Gaussian distribution. The noise in a particular direction $i$ 
is temporally correlated through a correlation function $C_i$, while being
uncorrelated with other directions $j\neq i$:

\begin{eqnarray}
P(\vec{r}^{\prime}) &=& \int{\cal{D}}{\roarrow{f}}P[\vec{f}] \delta \left(\vec{r}^{\prime} -\vec{r}(t)\right)\cr
P[\vec{f}] &=& N\exp{\left[ \displaystyle - \int ds
\int ds^\prime \sum_{ij} {{f_i(s)f_j(s^\prime)}\over{2\langle
f_i(s)f_j(s^\prime)\rangle }} \right]}\cr
\langle f_i\rangle &=& 0; \hskip 0.5 cm \langle f_i(t)f_j
(t^{\prime})\rangle = \displaystyle{ {{D_i}\over{\tau^c_i}}C_i
\left( {{|t-t^{\prime}|}\over{\tau^c_i}} \right)}\delta_{ij}. 
\label{e2}
\end{eqnarray}
$N$ is a normalization constant for $P$, and the subscripts $i$, $j$
correspond to different spatial components of the multi-dimensional vectors. 

The above representation of the probability distribution allows us to
write down the equation of motion for the probability distribution
$P(\vec{r})$ which implicitly depends on time through its coordinate
and the Langevin equation (\ref{e1}). The equation takes the
form of a continuity equation $\dot{P} = -\vec{\nabla}\cdot\vec{J}$. In
the weak noise limit, the correlation time $\tau^c_i$ is much smaller
than the diffusion time $\tau_i^D \equiv L_i^2/D_i$ over one period
$L_i$ of the potential, and the time $\tau_i^\gamma \equiv L^2_i/U_0$ for
an overdamped particle to fall from a potential $U_0$. Under these conditions, 
the current density $\vec{J}$ satisfies the Fokker-Planck form \onlinecite{r8}:

\begin{eqnarray}
\displaystyle J_i &=& W_iP - {{\partial}\over{\partial r_i}}\left[\Theta_iP\right]\nonumber\\
\displaystyle \Theta_i &=& D_i\left[1+\mu^i_1\tau^c_i {\bf{M}}_{ii} -
{{(\tau^c_i)^2}\over{2}}\mu^i_2({\bf{R}} - {\bf{M^2}})_{ii} + 
O(\tau^c_i)^3\right]
\label{e4}
\end{eqnarray}
where the matrix elements ${\bf{M}}_{ij} \equiv \partial
W_i/\partial r_j$, ${\bf{R}}_{ij} \equiv
\sum_kW_k\partial^2W_i/\partial r_j\partial r_k$, and $\mu^i_1$ and 
$\mu^i_2$ are the first and second moments respectively of the correlation
function $C_i$  \onlinecite{r8}. The role of color is thus to make the effective diffusion
coefficient $\Theta_i$ position-dependent in a well-defined manner (i.e., 
a function only on the potential $U(\vec{r})$). 

We solve Eq. (\ref{e4}) for $\vec{J}$ in steady-state ($\dot{P} =
0$)  and impose periodicity on the  
probability density $P$. In 
conjunction with the periodicity for $U$, this gives us a set of integral
equations for $\vec{J}$. In particular in two dimensions, these read:
\begin{eqnarray}
\displaystyle \int^{L_x}_0dxJ_x(x,y)e^{-\phi_x(x,y)}  &=& P(0,y)\Theta_x(0,y)
[1-e^{-\phi_x(L_x,y)}]\cr
\displaystyle \int^{L_y}_0dyJ_y(x,y)e^{-\phi_y(x,y)}  &=& P(x,0)\Theta_y(x,0)
[1-e^{-\phi_y(x,L_y)}]\cr
\displaystyle \phi_x(x,y) &=& \int^x_0dz{{W_x(z,y)}\over{\Theta_x(z,y)}}\cr
\displaystyle \phi_y(x,y) &=& \int^y_0dz{{W_y(x,z)}\over{\Theta_y(x,z)}}
\label{e7}
\end{eqnarray}
where $L_{x,y}$ are the periods along $x$ and $y$ directions 
respectively, and the coordinate zero is an arbitrary reference point on
the $x-y$ plane. 

Simplifying Eq. (\ref{e7}) elucidates the role of the correlation and
the asymmetry terms, which sit on the right hand side. Expanding the square 
brackets to the right to the first significant order in $\tau^c_i (\tau^c_1
\equiv \tau^c_x, \tau^c_2 \equiv \tau^c_y)$ yields $[1-e^{-\phi}]
\approx \phi$, where $\phi$ is given by:
\begin{eqnarray}
\displaystyle{\phi_x(L_x,y) = -{{(\tau^c_x)^2}\over{D_x}}\int^{L_x}_0dxW_x
(x,y)\left[{{\mu_2^x}\over{2}}\left({{\partial W_x}
\over{\partial y}}\right)^2 \right. }\cr
\displaystyle{\left. - \left({{3\mu_2^x}\over{4}} - {{{\mu_1^x}^2}\over
{2}}\right)W_x{{\partial^2W_x}\over{\partial x^2}} - {{\mu_2^x}\over{2}}
W_y{{\partial^2W_x}\over{\partial x\partial y}} \right]}
\label{e8}
\end{eqnarray}
and an analogous equation for $\phi_y(x,L_y)$.
This term is non-vanishing as long as the noise is {\it{correlated}}
($\tau^c_i\neq 0$) and the potential in two dimensions is {\it{asymmetric}},
 i.e., the integral over one period (Eq. \ref{e8}) is non-zero even
though the integrand itself is periodic \onlinecite{rMillonas}. 

The origin of rotation can now be traced to the existence of color and
asymmetry, which makes the integral in (\ref{e8})  and thus the right-hand side of Eq.
(\ref{e7}) non-zero. This prevents $\vec{J}$ from being identically zero or constant. 
Now, in steady-state the divergence of $\vec{J}$ is zero, so the only
way for $\vec{J}$ to not be identically zero within periodic boundary
conditions is to have the curl of $\vec{J}$ not identically zero.
{\it{This necessitates the current density to have local rotational fluxes.}}
On the other hand, if we take the limit of white noise ($\tau_{i}^c = 0$,
 $\forall$ $i$) or
make the integral in Eq. (\ref{e8}) vanish (symmetric potential), the
right-hand side of Eq. (\ref{e7}) is zero, and then $\vec{J}$ can
have zero curl. In fact, we can make the statement stronger: for
white noise, the effective diffusion constant $\Theta_i = D_i$ 
is position-independent. In that case, the Fokker-Planck equation and
the steady-state lead to the following set of equations:
\begin{eqnarray}
\vec{\nabla}&\cdot&\vec{J} = 0 \cr
\vec{\nabla}&\times&(\vec{J}e^{U(\vec{r})/D}) = 0.
\label{e9}
\end{eqnarray}
These equations, with periodic boundary conditions and detailed balance, lead 
to $\vec{J}\equiv 0$ \onlinecite{r6} which corresponds to an equilibrium
Maxwell-Boltzmann distribution for the probability density $P\sim \exp{[-U/D]}$.
In other words, in white noise in steady-state, there is no current density 
whatsoever. The damping in Eq. (\ref{e1})
breaks time-reversal symmetry in the system,
while color breaks detailed balance, producing rotations. 

Having thus established the necessary existence of rotational fluxes in a
colored noise, we now move on to a concrete example for an 
asymmetric potential in two dimensions as shown in Fig. \ref{f2}. The
parameter $a$ is the measure of the potential asymmetry. 
This potential is identical with respect to the transformation $x \leftrightarrow y$. We choose 
our correlation function for both the $x$ and $y$ variables to be decaying 
exponentials in time, and assume in addition $\tau^c_i\equiv\tau^c$ and $D_i\equiv D$ to remove 
any superficial differences between the $x$ and $y$ directions. 

We fix the boundary conditions of our potential by fixing $P(0,y) =
P(x,0)$ = constant for the sake of definiteness. For a particular 
instantiation of our general arguments for non-zero rotation, we make an 
ansatz at this stage: we assume that $J_x$ is a
function of $y$ only and $J_y$ is a function of $x$. This ansatz is
consistent with the form of Eq. (\ref{e7}) and trivially satisfies
the steady-state condition $\vec{\nabla}\cdot\vec{J} = 0$. 
Then  $J_x(y)$ is proportional to $\phi_x(L_x,y)$
of Eq. (\ref{e8}) divided by $\int^{L_x}_0dx\exp{[-\phi_x(x,y)]}$,
with an analogous equation for $J_y(x)$.
For the potential in Fig. \ref{f2}, Eq. (\ref{e8}) simplifies to
$\displaystyle{\phi_x(L_x,y) = -{{3(\tau^c)^2}}U_0^3a\pi[4\sin{2y} + \sin{4y}]}/4D$.

Figure \ref{f3} shows the resulting fieldplot for the current density $\vec{J}$ 
over one period of the potential. One immediately sees local rotational
fluxes separated by saddle points. There are drifts along $x$ for fixed $y$ 
which change sign as described schematically in Fig. \ref{f1} and produce a 
local rotational flux.  The role of {\it
{spatial asymmetry}} and {\it{ temporal correlation}} is clear by inspection 
of the expression for $\phi_x(L_x,y)$;  the current density $\vec{J}$ is zero if we set either 
$\tau^c$ or $a$ to be zero. In 
other words, the correlated noise breaks detailed balance, thereby exploiting
 the spatial asymmetries in the potential to produce local drifts and 
rotations. 

The relation between the potential profile in Fig. \ref{f2} and the flow
pattern in Fig. \ref{f3} can be summarized as follows: for a fixed $x$
($y$) coordinate, the sense of the drift along $y$ ($x$) is given by the 
asymmetry integral in Eq. (\ref{e8}). Note that the rotations take the 
particle in and out of potential hills and valleys, implying that energy is 
not conserved in the rotation process. The particle does not follow any of the
equipotentials, neither does the direction of the rotation depend on the
initial conditions of the system. The noise-driven rotation is a nonequilibrium
process which disobeys the fluctuation-dissipation theorem, and the sense of 
the rotation is determined entirely by the moments of the correlation function 
and potential asymmetries. 

Varying the potential leads to a whole 
range of steady-state flow patterns, including laminar flow (global
drift), global rotation (over one unit cell of the potential), and their
various combinations. To get a drift in any one coordinate, for
example, we need an asymmetric potential (such as a ratchet) in that 
direction. Combining asymmetric potentials in various ways, we can generate the
following classes of noise-driven fluxes \onlinecite{r8}:
\bigskip
\bigskip
\bigskip
\bigskip
\bigskip
\bigskip

\begin{quasitable}

\begin{tabular}{cc}
\tableline
\\
Flow pattern& Potential Form\\
\\
\tableline
\\
Rotation & Ratchets coupled in $x$ and $y$ \\
\\
Laminar flow & Decoupled ratchets in $x$ and $y$\\
\\
Rotation + net drift & Coupled ratchets asymmetric under\\
 & $x \leftrightarrow -x$, $y \leftrightarrow -y$\\
\\
\tableline
\end{tabular}
\label{table1}
\bigskip

\end{quasitable}
Our arguments are generalizable to higher dimensions; 
breaking other kinds of symmetries in $N$-dimensions will lead to a 
complex network of flow patterns. 

The broken detailed balance and spatial asymmetry essential
for current generation could be introduced in a variety of ways;
in our case, we consider correlated noise and asymmetry in the potential.
One could obtain similar results by introducing both spatial asymmetry and
temporal correlation in the potential itself, leading to currents in a 
``flashing potential'' \onlinecite{r2}. On the other hand, one could include 
both of these in
the noise itself and leave the potential symmetric with respect to $x$
and $y$ individually. This can be
accomplished, for example, by a noise that has more kicks on average in one
direction than in the other, thereby incorporating both broken spatial symmetry 
and detailed balance \onlinecite{r9}. 

One-dimensional ratchets have been explored under a variety of experimental conditions
\onlinecite{r10,r102,r11}. To observe noise-induced rotation in higher dimensions, we now propose
a realistic experimental set-up. 
Consider a system of 0.07 - 0.1 $\mu$m charged, fluorescent 
polystyrene beads suspended in an aqueous solution at room temperature in 
a two-dimensional potential. The periods of the potential are chosen to be $L_x = L_y
\approx 1$ $\mu$m, with an asymmetry fraction $\sim$
0.4 and maximum energy $U_0 \approx$ 75 meV. A set of crisscrossing 
electrodes is lithographically patterned to generate the two-dimensional 
ratchet potential similar to the one-dimensional potential in \onlinecite{r10}. 
Alternatively we can build a ratchet optically as in Ref. \onlinecite{r102}, for example, 
by passing a low intensity laser light
through a patterned reticle. For a colored noise with a 10-40 Hz
bandwidth generated electronically or optically, the particles will
settle into slow circular orbits of $\sim 1$ $\mu$m diameter. Our
calculations give us an estimated period of rotation of 1-10 hours
\onlinecite{r8}. Such fluorescent vortex patterns should be observable using a microscope.

We have analyzed the steady-state dynamics of an overdamped classical particle in an 
arbitrary multi-dimensional potential driven by a noise with an
arbitrary correlation function. For non-zero temporal correlations and
asymmetries in the potential, current production in terms of
rotations and drifts is expected. We have demonstrated rotation explicitly in
two dimensions in the limit of small correlation times. Such rotations
could be prototypes of periodic nonequilibrium processes such as molecular 
engines. By suitably tailoring the potential, one can 
generate a host of nonequilibrium flow patterns of the particle, including global 
drifts and rotations in combination. We have proposed a way of monitoring rotations in 
a realistic experimental set-up.

\acknowledgements We would like to thank O. Pierre-Louis for suggesting
the problem, and C. Jayaprakash and F. J${\ddot{\rm{u}}}$licher for
useful discussions. AWG thanks the Ohio State University Presidential
Fellowship, and SVK thanks the Department of Energy - Basic
Energy Sciences and Division of Materials Sciences for support.

\begin{figure}
\caption{Schematic description of rotation over one unit cell of a 
two-dimensional periodic potential, caused by spatial asymmetry and temporal 
correlations. One-dimensional drifts are produced by asymmetric potentials in 
$x$ and $y$, in conjunction with correlated noise. The drifts switch 
directions owing to coordinate-dependent changes in overall sign of the 
potential asymmetry, and together produce rotation.} 
\label{f1}
\end{figure}

\begin{figure}
\caption{Two-dimensional contour plot of a potential with translational 
asymmetry in $x$ and $y$. The potential chosen is given by
$U(x,y) = U_0[\sin{x}\sin{y} - a\sin{2x}\sin{2y}]$ with $U_0 = a = 1$. Dark areas correspond
to valleys and bright patches correspond to hills in the potential
landscape.}
\label{f2}
\end{figure}

\begin{figure}
\caption{Local rotations ($\vec{J}$) produced by drifts arising out of
translational asymmetries in the potential in Fig. \ref{f2} and
correlation in the noise.  There is no global rotation, and the
steady-state flow consists only of cycles and saddle points. At fixed
values of $x$ there are global drifts in $y$ that reverse sign so as to
generate the rotations. The arrow lengths are proportional to
$(\tau^c)^2U_0^3a/D$, so that as $a$ or $\tau^c \rightarrow 0$, the arrow
length shrinks to zero and the local current densities vanish.}
\label{f3}
\end{figure}


\begin{thebibliography}{16}
\bibitem[*]{byline} E-mail: avik@campbell.mps.ohio-state.edu
\bibitem[\dag]{byline} E-mail: khare@pacific.mps.ohio-state.edu

\bibitem{rAst3} R. D. Astumian, P. B. Chock, T. Y. Tsong, Y. D. Chen and H. V.
Westerhoff, Proc. Natl. Acad. Sci. U.S.A. {\bf{84}}, 434 (1987). 

\bibitem{r2} A. Ajdari and J. Prost, C. R. Acad. Sci. Paris {\bf{315}},
1635 (1992). 

\bibitem{r1} M. Magnasco, Phys. Rev. Lett. {\bf{71}}, 1477 (1993). 

\bibitem{rAst} R. D. Astumian and M. Bier, Phys. Rev. Lett. {\bf{72}},
1766 (1994). 

\bibitem{dhr94} C. R. Doering, W. Horsthemke, and J. Riordan, Phys. Rev. Lett.
{\bf 72}, 2984 (1994).

\bibitem{rLeib} S. Leibler, Nature (London) {\bf{370}}, 412 (1994). 

\bibitem{rAst2} R. D. Astumian, Science {\bf{276}}, 917 (1997). 

\bibitem{rM} M. Magnasco, J. Stat. Phys. {\bf{93}}, 615 (1998). 

\bibitem{r10} J. Rousslet {\it{et al.}}, Nature (London) {\bf{370}}, 446
(1994).

\bibitem{r102} L. P. Faucheaux {\it{et al.}}, Phys.  Rev. Lett. {\bf{74}}, 1504
(1995). 

\bibitem{r11} C. Mennerat-Robilliard {\it{et al.}}, Phys. Rev. Lett. {\bf{82}}, 851 (1999). 

\bibitem{r12} H. Linke, W. Sheng, A. Lofgren, H. Q. Xu, P. Omling and P.
E. Lindelof, Europhys. Lett. {\bf{44}}, 341 (1998). 

\bibitem{r13} I. Zapata, R. Bartussek, F. Sols and P.
H${\ddot{\rm{a}}}$nggi, Phys. Rev. Lett. {\bf{77}}, 2292 (1996). 

\bibitem{r3} F. J${\ddot{\rm{u}}}$licher, A. Ajdari and J. Prost, Rev.
Mod. Phys. {\bf{69}}, 1269 (1997).

\bibitem{r6} H. Risken, {\it{The Fokker-Planck Equation}} (Springer-
Verlag, New York, 1989). 

\bibitem{r7} R. F. Fox, Phys. Rev. A {\bf{33}}, 467 (1986). 

\bibitem{r8} A. W. Ghosh {\it{et al.}}, {\it{in
preparation}}. 

\bibitem{rMillonas} M. M. Millonas and M. I. Dykman, Phys. Lett. A
{\bf{185}}, 65 (1994). 

\bibitem{r9} R. Lahiri, LANL preprint archive, cond-mat/9607099.

\end{thebibliography}
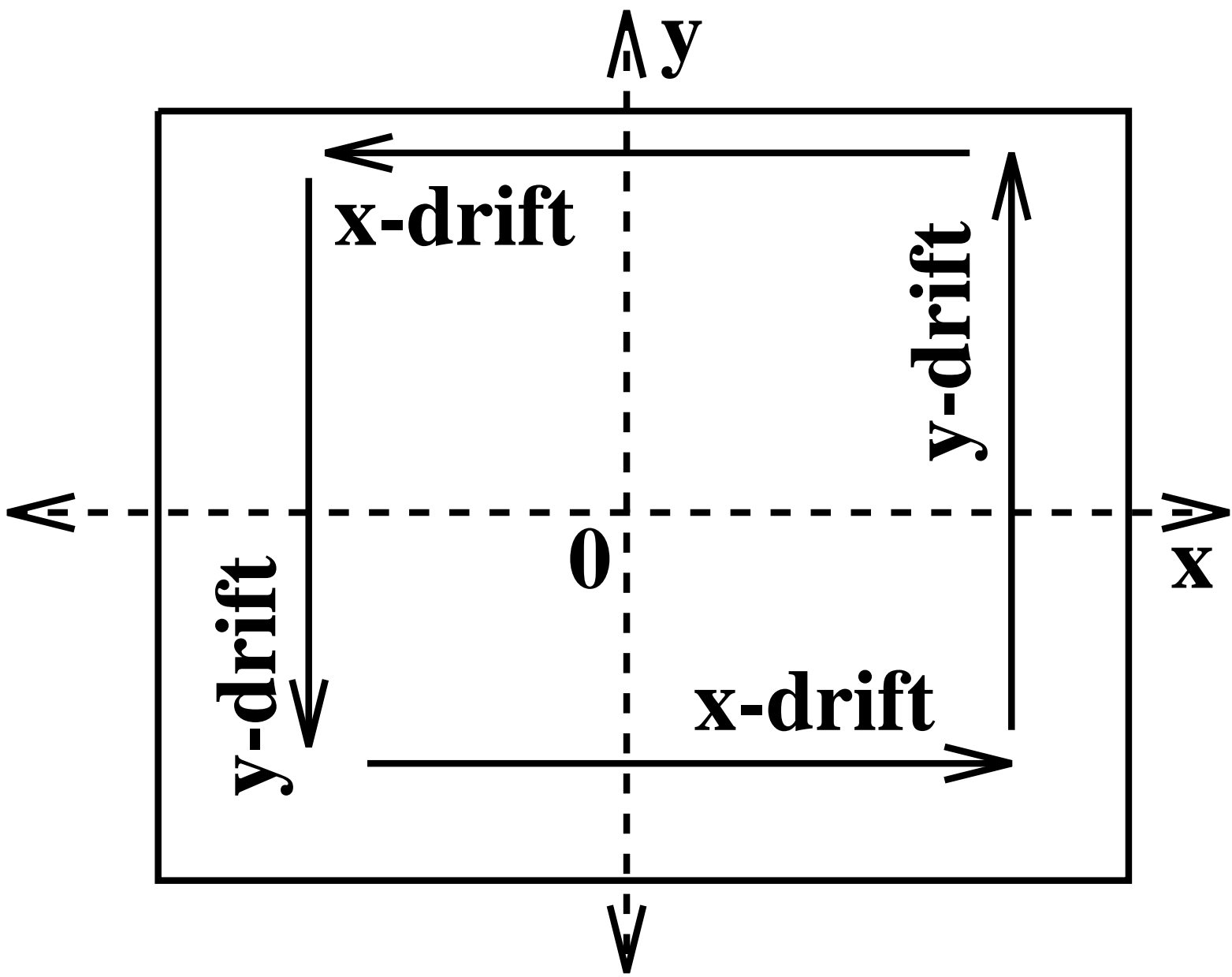
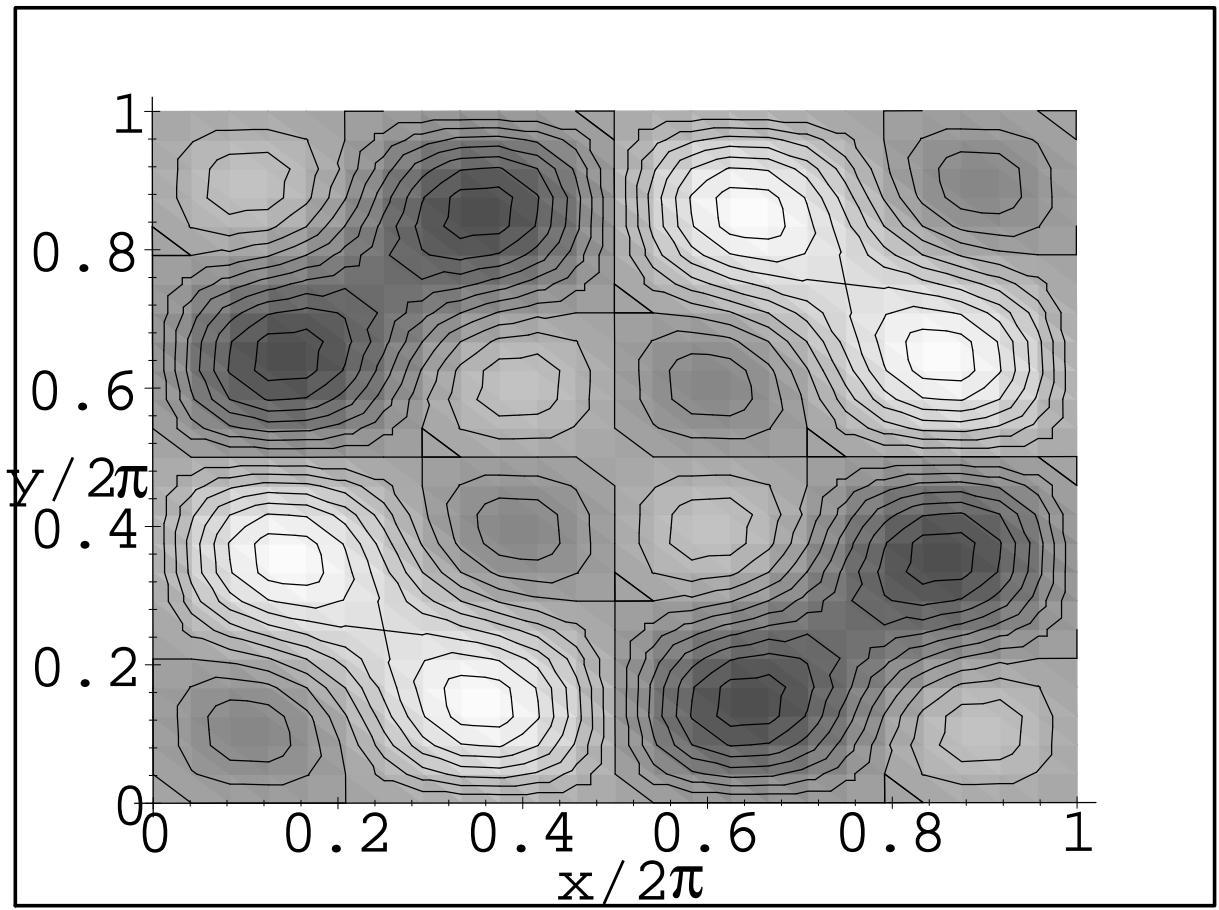
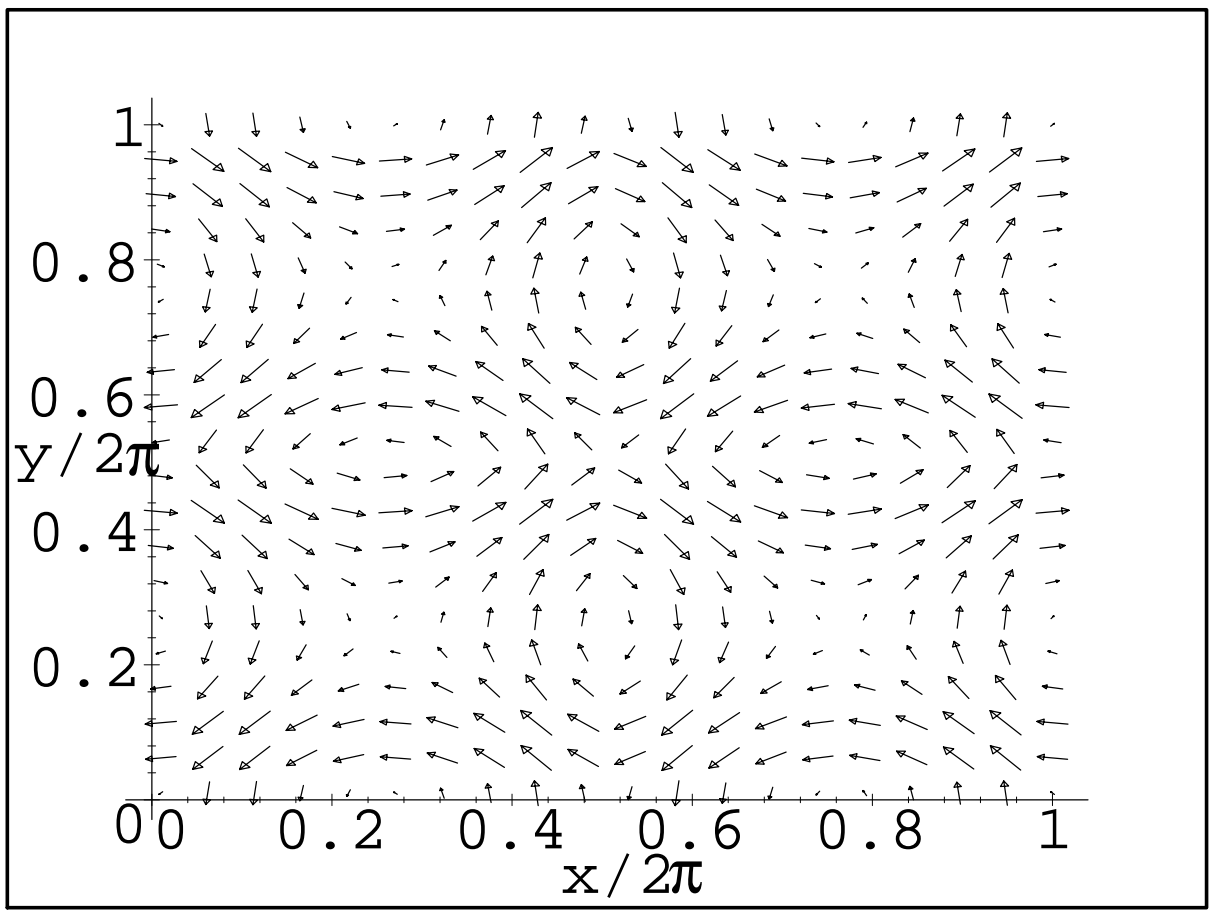
\end{document}